\documentclass[aps,prl,twocolumn,amsmath,amssymb]{revtex4}
\usepackage{graphicx}
\usepackage{dcolumn}
\usepackage{bm}

\begin{document}

\title{Conductance Quantization in Graphene Nanoribbons}

\author{Yu-Ming Lin}
\email{yming@us.ibm.com}
\author{Vasili Perebeinos}\author{Zhihong Chen}
\author{Phaedon Avouris}
\affiliation{IBM T.\ J.\ Watson Research Center, Yorktown Heights,
NY 10598, USA}

\begin{abstract}
We report the experimental observation of conductance quantization
in graphene nanoribbons, where 1D transport subbands are formed  due
to the lateral quantum confinement. We show that this quantization
in graphene nanoribbons can be observed at temperatures as high as
80\,K and  channel lengths as long as 1.7\,$\mu$m. The observed
quantization is in agreement with that predicted by theoretical
calculations.
\end{abstract}

\maketitle

The isolation of graphene\,\cite{novoselov_nature2005}, a single
sheet of graphite, has incited numerous studies because of
fundamental physics interests\cite{rise_graphene,Neto_review_MPR}
and promising applications for carbon-based
electronics\cite{carbon_electronics_review}. As a truly
two-dimensional system and a zero-gap semiconductor where the
electrons and holes behave as massless fermions, graphene possesses
distinctly different transport properties from that of conventional
2D and 3D electronic materials. However, in order to utilize their
remarkable electrical characteristics in certain applications, it
would be highly desirable to produce a band gap in graphene, and
therefore, intense efforts are being made to explore the properties
of low-dimensional (1D and 0D) graphene
nanostructures\cite{Han_PRL2007,zhihong_graphene2007,CB_inGraphene_APL,CB_in_GNR_PRL}.

By patterning graphene into a narrow ribbon structure, the carriers
are laterally confined  to form a quasi-one-dimensional (1D) system,
similar to the case of carbon nanotubes. Due to the linear
dispersion relation $E= {\rm v_f}\hbar k$ and high Fermi velocity
(${\rm v_f}\sim 10^6$m/s) in graphene, the quantization energy of
graphene nanoribbons (GNRs) can be substantially larger than that of
conventional semiconducting materials of the same dimension and
parabolic dispersion. The formation of 1D subbands in GNRs is
expected to yield an energy gap for certain ribbon widths and
crystallographic directions\,\cite{dresselaus_graphene_edge1996}. In
terms of transport, this quantum confinement can also lead to
quantized conductance, which is one of the most important transport
characteristics of mesoscopic physics.

Recently, experimental studies on GNRs have revealed a
thermally-activated
conductivity\,\cite{zhihong_graphene2007,Han_PRL2007}, suggesting
the presence of a size-dependent energy gap. However, there remains
some controversy in understanding the transport behavior in
realistic GNR devices, where the imperfect and unknown edge
configurations may lead to Columbic-blockade type transport at low
temperatures\,\cite{CB_in_GNR_PRL}. Furthermore, an energy gap could
also be induced by substrate
interactions\cite{BG_graphene_substrate}. Despite numerous
theoretical predictions on
GNRs\cite{conductance_quantization_PRB06,SemiconductingGNR_APL_white,Electronic_struc_in_GNR_PRB_Brey},
conductance quantization has yet to be reported experimentally.

In this Letter, we present electrical transport measurements for GNR
devices with lateral widths of 30\,nm, and report conductance
plateau features in GNRs at temperatures as high as 80\,K. By
modulating the Fermi energy by a back gate, we observe conductance
quantization plateaus for electron and holes in the same device,
both possessing comparable transmission probability for each 1D
conduction mode. Our results provide the direct experimental
evidence of quantum size confinement effects and the formation of
subbands for 1D graphene nanostructures.

\begin{figure}
\vspace{0em}
\includegraphics[width=8cm]{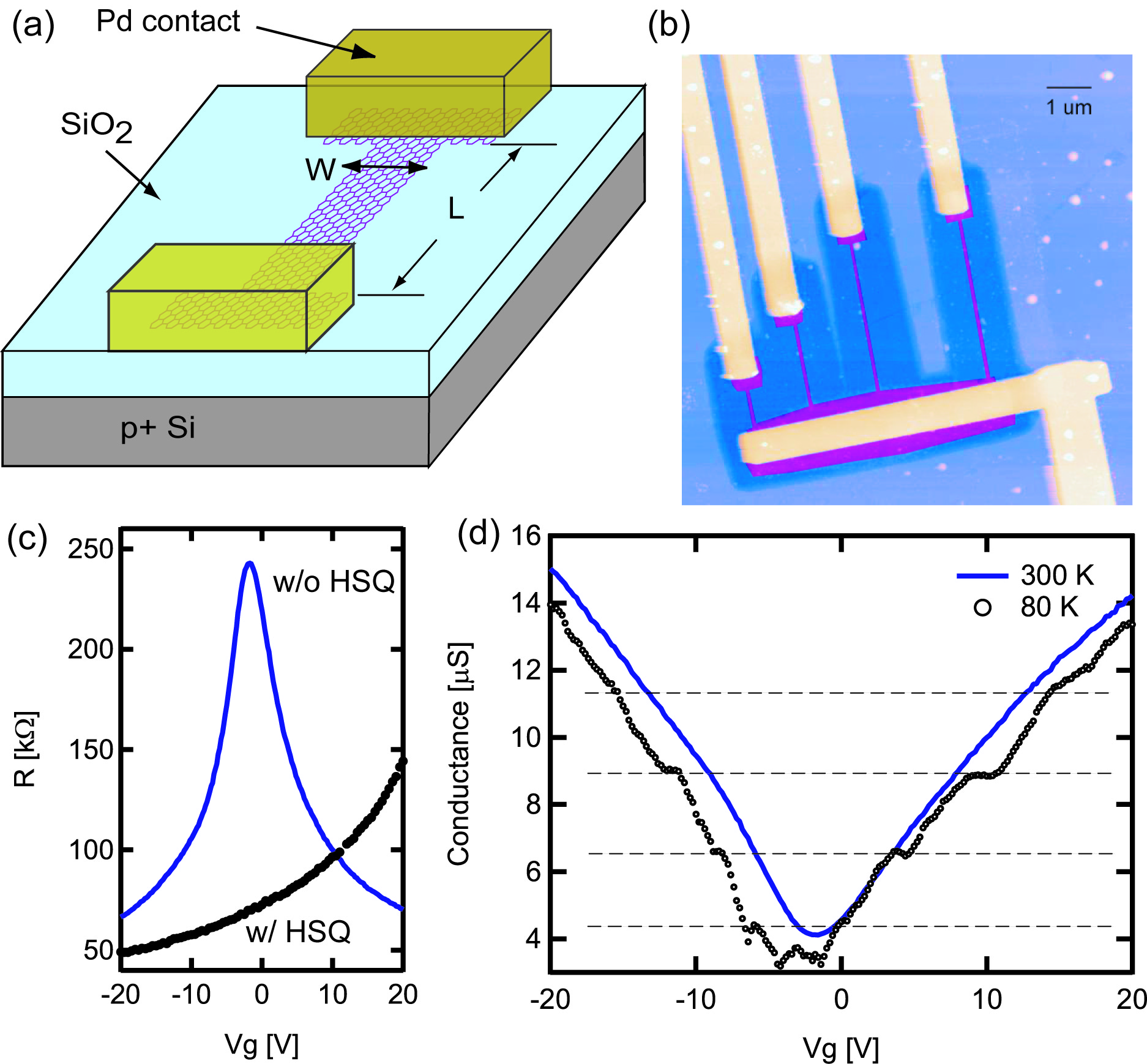}
\caption{(a) Schematics of the GNR device, where the graphene is
contacted by two Pd metal leads, and the $p$-doped Si substrate acts
as the gate electrode. (b) AFM image of GNR devices with different
channel lengths. (c) Resistance of a GNR device measured as a
function of gate voltage before and after the HSQ layer is removed,
showing the impact of HSQ on electrical behaviors. Both measurements
were performed in vacuum after annealing at 135$^\circ$C. (d)
Conductance of a GNR device ($W=30$\,nm and $L=850$\,nm) measured as
a function of gate voltage at 300\,K and 80\,K. The bias voltage is
10\,mV. } \label{Fig1}\vspace{-1em}
\end{figure}

The GNR devices in our study, schematically shown in
Fig.\,\ref{Fig1}(a), were fabricated from mechanically-exfoliated
graphene sheets on a $p$-doped Si substrate covered with 300-nm
thick SiO$_2$. Raman spectroscopy and atomic force microscope (AFM)
measurements were employed to identify single-layer graphene. For
the details of the device fabrication, we refer to
Ref.\,\cite{zhihong_graphene2007,Lin_graphene_noise}. The source and
drain contacts are made of Pd, and the Si substrate acts as the back
gate. Narrow ribbons of graphene are formed by oxygen plasma RIE
etching using a patterned HSQ (hydrogen silsesquioxane) layer as the
protective mask. Fig.\,\ref{Fig1}(b) shows the AFM image of an array
of 30-nm-wide GNRs after this HSQ layer is removed in hydrofluoric
acid (HF) solution. We found that electrical properties of the GNRs
are strongly affected by this HSQ layer, which needs to be removed
in order to reveal their intrinsic transport properties and, more
importantly, the conductance quantization behaviors at low
temperatures, as discussed below. Fig.\,\ref{Fig1}(c) shows the
resistance as a function of gate voltage $V_g$ of a GNR device
before and after the HSQ layer is removed. We note that the presence
of HSQ leads to a significant positive shift of the Dirac point
voltage and a reduction of carrier mobility in graphene. In
contrast, after removing the HSQ layer, the GNR exhibits ambipolar
behavior with the resistance maximum associated with the Dirac point
occurring at $V_g\simeq$ 0\,V, indicating a negligible unintentional
doping in the final device.

Transport measurements of GNRs were performed in vacuum
($\sim10^{-7}$\,torr), and the devices studied all possess a Dirac
voltage near $V_g\sim$ 0\,V after annealing at 135$^\circ$C.
Fig.\,\ref{Fig1}(d) shows the conductance $G$ of a GNR device
($L=850$\,nm and $W=30\,$nm) measured as a function of gate voltage
$V_g$ at a dc bias of 10\,mV. At room temperature, the conductance
curve $G(V_g)$ in Fig.\,\ref{Fig1}(d) resembles that of bulk
graphene\,\cite{rise_graphene}, showing the characteristic "V" shape
that reflects symmetric hole and electron transports at negative and
positive gate voltages, respectively. While the overall conductance
curve $G(V_g)$ of the GNR displays little variation as temperature
decreases, several plateau features start to appear in the measured
$G(V_g)$ curve and become apparent for $T< 100$\,K (see
Fig.\,\ref{Fig1}(d)). The slight asymmetry in the slope of the $n$
and $p$ branches is likely associated with the gate oxide
hysteresis. Nevertheless, we note that these conductance plateaus
are observed in both electron and hole branches with nearly the same
conductance values and an equal spacing (see dashed line in
Fig.\,\ref{Fig1}(d)).

\begin{figure}
\vspace{1em}
\includegraphics[width=7cm]{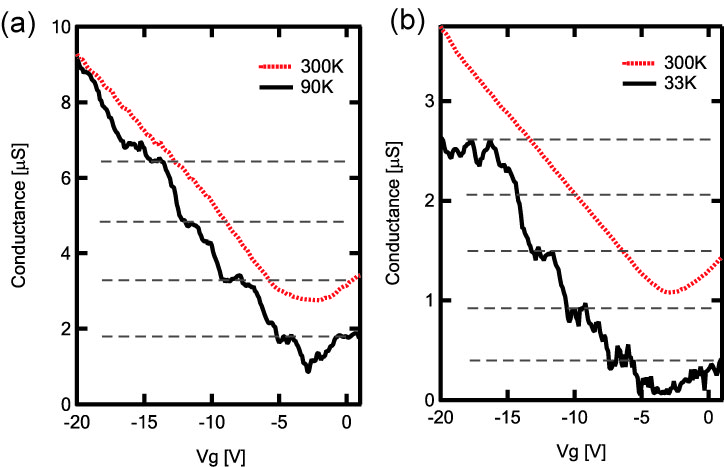}
\caption{Conductance measured as a function of gate voltage for
30-nm-wide GNR devices with different channel lengths. The GNR
channel lengths are 900\,nm and 1.7$\mu$m for (a) and (b),
respectively. At low temperatures, both devices show equally-spaced
conductance plateaus with $\Delta G\simeq$ 1.7\,$\mu$S and
0.6\,$\mu$S for (a) and (b), respectively. The bias voltage is kept
at 10\,mV.} \label{fig2}\vspace{-1em}
\end{figure}
Conductance plateau features similar to those found in
Fig.\,\ref{Fig1}(d) are also observed in other 30-nm-wide GNR
devices for channel lengths $L$ up to 1.7\,$\mu$m.
Figs.\,\ref{fig2}(a) and (b) show the $G(V_g)$ measured for two GNRs
with $L=900$\,nm and 1.7\,$\mu$m, respectively. At low temperatures,
both GNRs exhibit equally-spaced conductance steps as $V_g$ varies.
From Fig.\,\ref{Fig1}(d) and Fig.\,\ref{fig2}, while the spacing
$\Delta G$ between conductance plateaus depends on the channel
length, where $\Delta G$ decreases with increasing $L$, the number
of plateaus is roughly constant in the same gate voltage span ($\sim
20$\,V) for these GNRs of the same width. It is also important to
note that these conductance plateau structures are highly
reproducible under different thermal cycles, and they generally
become more well-defined as $T$ decreases, in particular for longer
channel devices. However, as the temperature further decreases below
10\,K, the $G(V_g)$ traces of these GNRs are usually overwhelmed by
a pronounced fluctuation background as a function of gate voltage.
Unlike the plateau structures, the low-$T$ background fluctuations
are not retraceable in different thermal cycles, indicating that the
low-$T$ background fluctuations may be due to the universal
conductance fluctuation (UCF)
phenomena\,\cite{zhihong_graphene2007}.

Based on these experimental findings, we attribute the conductance
plateaus observed in GNRs to the formation of 1D conduction modes in
nanoribbons. There have been numerous theoretical studies on the GNR
electronic structure and its dependence on the ribbon width and the
cutting angle using $\pi$-orbital tight-binding
models\cite{SemiconductingGNR_APL_white,conductance_quantization_PRB06}
and the first principle
calculations\cite{GNR_Louie,Electronic_struc_in_GNR_PRB_Brey}. To a
first order approximation, the band structure of GNRs can be
described by zone folding the graphene band structure. In a graphene
nanoribbon of width $W$, the wave vector perpendicular to the
transport direction has a quantization requirement $k_{\perp}W=\pi
m$, giving rise to various 1D subbands, each with the dispersion
relation given by
\begin{equation}
E_m(k_{\parallel}) = \pm \hbar {\rm v_f}\sqrt{k_{\parallel}^2+(m+\alpha)^2\pi^2/W^2},
\end{equation}
where $m=0,\pm 1,\pm 2, ...$ is an integer for the subband index and
$k_{\perp}$ and $k_{\parallel}$ are the wave-vector perpendicular
and parallel to the transport direction, respectively. Here
$0\le|\alpha|<0.5$ depends on the crystallographic orientation of
the GNR, and yields a bandgap $\Delta_{GNR}=2\Delta E|\alpha|$,
where $\Delta E = \hbar {\rm v_f}\pi/W$ is the energy separation
between the 1D subbands. We note that in GNRs, the wavefunction has
to vanish at the ribbon edges, in contrast to the periodic boundary
condition in the case of carbon nanotubes (CNTs). This leads to a
different quantization requirement in CNTs, namely $k_{\perp}W_{\rm
CNT}=2\pi m$, and as a result twice as large energy separation in
CNTs for the same circumference length $W$. In addition, the 1D
subbands of GNRs are singly degenerate as opposed to those of
CNTs\cite{White_NL_2007}, where the orbital degeneracy associated
with the K-K' bands is lifted with the splitting determined by the
GNR chirality.

For a more realistic model of the GNR band structure, we adopt the
tight-binding method with the next-neighbor hopping $t_{\rm
C}=2.7$\,eV. The Fermi velocity is related to the hopping integral
$t$ according to ${\rm v_f}=3at_{\rm C}/2\hbar\simeq10^6$\,m/s,
where $a\approx 0.246$ nm is a lattice constant of graphene. At the
ribbon edge, we assume enhanced hopping between the neighboring C
atoms by $12$\% to account for the C-C bond length
contraction~\cite{Louie_PRL06}. We find that in armchair GNRs
(unrolled zigzag CNTs), $\alpha\approx 0.27$, $0.4$, and $0.066$ in
the families $N=3p$, $N=3p+1$, $N=3p+2$ respectively, where $p$ is
an integer and $N$ is the number of dimer lines across the ribbon
width. As a result, the K-K' orbital splitting in families $N=3p+1$
and $N=3p+2$ is about $0.2\Delta E$ and $0.13\Delta E$,
respectively. On the other hand, for the $N=3p$ family, the subband
energies are separated by $0.46\Delta E$ (or $0.54\Delta E$), such
that the resulting subbands are roughly equally spaced with half the
quantization energy $\Delta E$.

Using the Landauer approach, the device conductance at a finite
temperature can be expressed by
\begin{equation}
G=\frac{2e^2}{h}\sum_i{\int{T_i(E)\left( -\frac{\partial
f_0}{\partial E}\right)}dE}, \label{eq:eq1}
\end{equation}
where $T_i(E)$ is the transmission probability of carriers in each
subband at energy $E$, and the factor 2 comes from the spin
degeneracy. At zero temperature, Eq.\,(\ref{eq:eq1}) can be reduced
to $G=2e^2/h \sum_i{t_i}$ where $t_i = T_i(E_f)$ and the summation
includes all the 1D modes below the Fermi energy. In our simulation,
we assume an energy-independent transmission probability $t$ for all
subbands.
\begin{figure}
\vspace{1em}
\includegraphics[width=9cm]{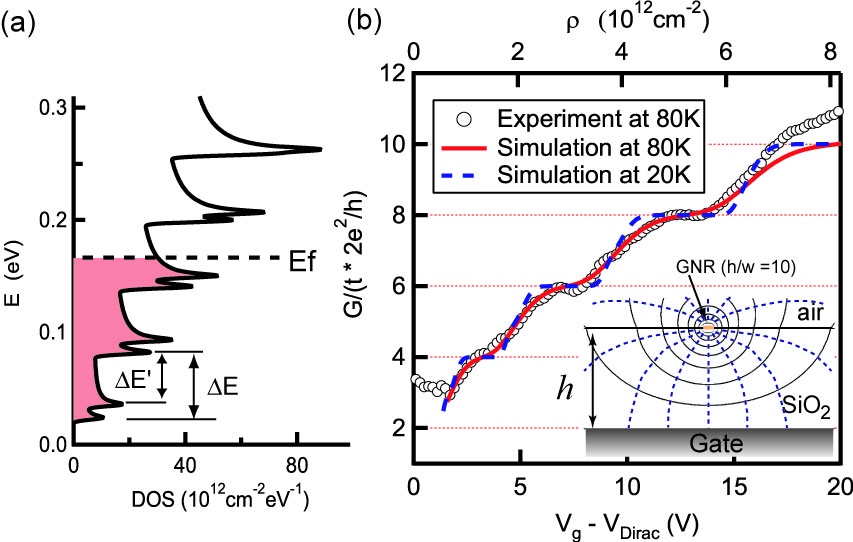}
\caption{(a) The density of states of a graphene nanoribbon.
(b)Calculated normalized conductance $G/(2te^2/h)$ as a function of
carrier density for graphene nanoribbon devices of $W=31$\,nm, using
armchair GNR with $N=253$. The inset illustrates schematically the
field lines and the equipotential surfaces of the GNR, demonstrating
a significant discrepancy from the case of a parallel-plate
capacitor.} \label{fig_simu}
\end{figure}

Now we compare simulations of the GNRs with the experimental
results. For a 30-nm-wide GNR as shown in Fig.\,\ref{Fig1}, we
estimate the quantization energy $\Delta E$ to be 60\,meV. From
simulations based on Eq.\,(\ref{eq:eq1}), we find that an energy
spacing of at least 6$kT$ is required in order to observe
conductance quantization features at a given temperature $T$. Since
the conductance steps are clearly visible in Fig.\,\ref{Fig1} at
80\,K, corresponding to a minimum energy separation of 40\,meV, the
energy splitting at the K-K' bands must be smaller than 20\,meV.
These facts can be accounted for, within the tight-binding model, by
adopting a (3$p$+1) family armchair ribbon to model the electronic
structure of the GNR in Fig.\,\ref{Fig1}. Fig.\,\ref{fig_simu}(a)
shows the calculated density of states of an armchair GNR with
$N=253$, corresponding to $W$=31\,nm. The quantization energy in
this ribbon is $\Delta E\approx 58$ meV and the separation between
the nearly degenerate pairs of bands is about $\Delta E'=0.8 \Delta
E\approx 46$ meV. Indeed, we find an excellent agreement between the
measurement and the simulation at 80\,K as shown in
Fig.~\ref{fig_simu}(b). In Fig.\,\ref{fig_simu}(b),  the step height
is $4e^2/ht$ because the steps associated with individual occupation
of nearly degenerate bands, which are separated by only 11\,meV,
cannot be resolved at this temperature. In order to further resolve
these nearly degenerate subbands, the measurements should be
performed below 10 K, the regime where universal conductance
fluctuation dominate the transport properties. From
Fig.\,\ref{fig_simu}(b), we calcuate the transmission probability
$t$ of this GNR to be 0.016. We note that while the transmission
probability may be a function of the length, energy and the contact
to the sample, the experimental results Fig.\,\ref{fig_simu}(b) can
be well-described by a constant $t$ at low energies. On the other
hand, the discrepancy between the simulated and measured conductance
of the fourth plateau around $V_g\ge17$\,V may be associated with
the enhanced transmission probability at high energies.

The carrier density in the GNR is related to the gate voltage as
$\rho=C_g(V_g-V_{\rm Dirac})$, where $C_g$ is a gate capacitance and
$V_{\rm Dirac}$ is a gate voltage corresponding to the Dirac point.
As shown by Fig.\,\ref{fig_simu}(b), we observe four conductance
plateaus over a $V_g$ span of 20\,V, corresponding to a shift of the
Fermi level by $E_{\rm f}\approx 260$ meV and a change in carrier
density of $8.4 \times 10^{12}$ cm$^{-2}$. This is comparable to the
carrier density in 2D graphene at the same Fermi energy $E_f$ given
by $\rho_{2D}=E_{\rm f}^2/\pi {\rm v_f^2}$. The dependence of $\rho$
on $V_g$ leads to an effective gate capacitance of $C_{\rm GNR}=
4.2\times10^{11}$\,cm$^{-2}$V$^{-1}$, which is significantly larger
than the value expected for 2D graphene, given by $C_{\rm 2D}\simeq
7.2\times10^{10}$\,cm$^{-2}$V$^{-1}$ for the gate dielectric of
SiO$_2$ with thickness $h=300$\,nm.  This difference is due to the
fact that when the aspect ratio of the GNR width to the oxide
thickness $(w/h)$ becomes much smaller than one, the field lines
between the GNR and the gate deviates from that of a parallel-plate
capacitor (see inset of Fig.\,\ref{fig_simu}(b)), resulting in an
increasing $C_{\rm GNR}/C_{\rm 2D}$ ratio as $(w/h)$ decreases. We
have numerically calculated the capacitance of the GNR of infinite
length $L$ in the back gate configuration where the SiO$_2$
dielectrics and the vacuum below and above the ribbon are explicitly
considered. For $h/w$=300/30, we find $C_{\rm GNR}/C_0\simeq$
10\footnote{The gate capacitance of GNRs can also be approximated by
the following semi-analytical approach. Assuming a homogenous
dielectric medium $\varepsilon$ below and above the GNR, the
capacitance due to the back gate can be expressed by
$\frac{C_0}{C}=\frac{2}{\pi}\arctan{\frac{W}{4h}}+\frac{W}{4h\pi}\ln{\left(1+\frac{16h^2}{W^2}\right)}$,
where $C_0=\varepsilon\epsilon/h$. The solution assumes a constant
charge density on the GNR. The gate capacitance of the device is
then approximated by $C_{\rm GNR}\simeq (C_0^{\rm air}+C_0^{\rm
SiO_2})/2$, yielding an enhancement factor of 9.}, which is within a
factor of 2 of the estimated values based on
experiments\footnote{The actual gate capacitance of the GNR is
expected to be smaller than the numerical simulations because of (i)
the non-ideal dielectrics used and (ii) the screening from source
and drain electrodes.}.

We also note that, since the device conductance at room temperature
is comparable to its low temperature value, the scattering mechanism
is not temperature sensitive in these GNRs, and therefore, the
dominate scattering process is mostly due to the impurity scattering
within the channel or the edge disorders. This results in a
transmission probability that decreases with the increasing channel
length. It also implies that these 1D transport channels are not
ideally ballistic. The fact that similar conductance quantization is
observed for both electrons and holes reflects the symmetry in the
electron and hole band structures as well as in their transmission
probability.

\begin{figure}
\includegraphics[width=8cm]{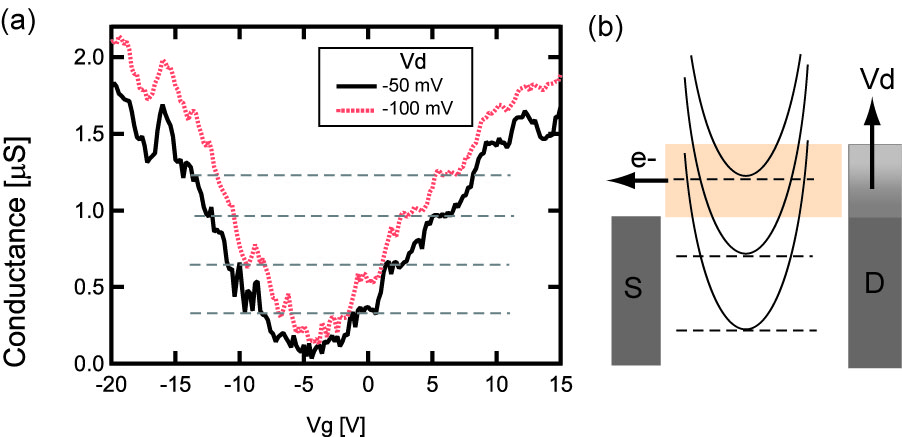}
\caption{(a) Conductance of a GNR device measured as a function of
gate voltage at $T$=15\,K for two drain biases, 50\,mV and 100\,mV.
The GNR device has a channel length and width of 1.7\,$\mu$m and
30\,nm, respectively. (b) Schematic energy band diagram of a GNR
with a drain bias voltage applied. The conductance is determined by
the number of 1D subbands accessible in the bias window. For a given
gate voltage, the number of 1D modes contributing to the transport
can be increased at a sufficiently large drain bias.} \label{Fig3}
\end{figure}

In addition to the Fermi level modulation by the gate control, the
number of 1D conduction modes that contribute to the transport in
GNRs  also depends on the bias voltage between the source and drain
contacts, yielding a $V_d$-dependent device conductance in
GNRs.\,\cite{conductance_quantization_PRB06} Fig.\,\ref{Fig3}(a)
shows measured conductance of a GNR device ($W=30$\,nm and
$L=1.7\,\mu$m) as a function of $V_g$ for two bias voltages $V_d$ of
50\,mV and 100\,mV at 15\,K, both exhibiting conductance plateaus
with the same quantization spacing $\Delta G$ as $V_g$ varies.
However, in Fig.\,\ref{Fig3}(a), it is noted that the conductance
curve $G(V_g)$ at $V_d=100\,$mV is always higher than that of
$V_d=50$\,mV by roughly one unit of quantized conductance $\Delta G$
for all gate voltages. In comparison, at $V_d=10$\,mV, this GNR
possesses a nearly identical $G(V_g)$ curve to that at $V_d=50$\,mV.
This dependence of conductance on the drain bias can be understood
through the schematic shown in Fig.\,\ref{Fig3}(b). As the
source-drain bias windows becomes larger than the energy spacing of
the 1D subbands, which is $\sim 67$\,meV for a 30-nm-wide GNR, the
number of 1D conduction modes is increased by one regardless of the
gate voltage, resulting in a higher conductance level by $\Delta G$.

In conclusion, we have fabricated graphene nanoribbon devices and
reported the first experimental observation of conductance
quantization phenomena in GNRs at temperatures as high as 80\,K and
channel lengths up to 1.7\,$\mu$m. We have performed temperature and
bias dependence studies, and shown that the energy spacing between
1D subbands of 30-nm-wide GNRs is around 50\,meV. The experimental
results are in excellent agreements with theoretical calculations
within the tight-binging approximation. The experimental findings
here provide an important step towards developing graphene-based
quantum devices.

The authors thank Bruce Ek for expert technical assistance.

\end{document}